\documentclass[preprint,12pt]{elsarticle}

\usepackage{graphicx}
\usepackage{amssymb}

\usepackage{lineno}

\journal{Physics Letters A}

\begin{document}

\begin{frontmatter}


\title{A proposal for detection of absolute rotation using superconductors and large voltages}



\author{E.M. Forgan}
\ead{E.M.forgan@bham.ac.uk}
\author{C.M. Muirhead,  A.I.M. Rae $\&$ C.C. Speake}

\address{School of Physics $\&$  Astronomy, University of Birmingham, B15 2TT, U.K.}

\begin{abstract}
We describe designs for practical detectors of absolute rotation, which rely on the creation of magnetic fields by charged objects that are rotating with respect to an inertial frame. Our designs, motivated by an original suggestion by R.M. Brady, utilize the properties of superconductors, both to shield and confine the magnetic fields, and also as the basis of a SQUID detector of the fields produced. We show that with commercially available SQUIDs, our designs can have sufficient sensitivity and signal-to-noise ratio to measure the sidereal rate of rotation of the Earth. We consider three different designs: two of these can also be configured to provide a confirmation of the form that Maxwell’s equations take in a rotating frame. We can also make a direct experimental test of whether low-frequency electromagnetic energy experiences the same inertial rest-frame as matter.
\end{abstract}


\end{frontmatter}


\section{Introduction}
\label{S:1}

Consider a long normal metal cylinder carrying a uniform excess electrical charge around its curved surface. If we now rotate the cylinder about its long axis, the charge will move with the metal, and constitute a circulating current resulting in a magnetic field parallel to the axis of the cylinder. If we can measure this field and know the distribution of charges, we can use it to deduce the rate of rotation. The experimental demonstration that physical movement of electrostatic charge has the same magnetic effect as an electric current was first conclusively demonstrated by Rowland \& Hutchinson's pioneering experiments~\cite{Rowland} in the 19$\rm th$ century, and were important in the development of understanding of electromagnetism. Here we use the same principle for a different purpose.  If we aligned our cylinder with the Earth’s axis then, with sufficient sensitivity, we could use it to measure the rate of rotation of the Earth relative to the inertial frame, which is presumably the rest of the Universe (Mach’s principle), without observing the stars. The idea of using superconductors and high voltages was tried by R.M. Brady in the 1980’s with the technology of the time using rotation rates $ \sim $ radians/sec. His first design\cite{Brady1-pub} gave signals that appear to be spurious, as they do not reverse on reversing the direction of rotation. The results from his later design were never published in full: Ref.~\cite{Brady2-online} is a brief account only available online, showing results limited by considerable drift and other complicating factors, but apparently of the magnitude expected from his and our predictions. We describe here three different designs, intended to avoid unwanted signals, and give a theoretical treatment of Brady's design in the Appendix.
 We find that it is realistic to construct an apparatus that can accurately measure the Earth’s sidereal rotation rate with a reasonable noise-integration time, so long as external magnetic fields are very low and remaining flux lines are well-pinned in the superconducting parts.
Two forms of the apparatus may also be used to check the form of Maxwell's equations in rotating frames by experiments using rotations with a shorter period.  The behaviour of the superconducting parts may at first appear counter-intuitive so we introduce the design concepts in a heuristic way
\section{Heuristic description of the design}
\label{heuristic}
In Figs.~$1~\&~2$ we show the evolution of a conceptual design. 1(a) shows the cross section of a positively charged long hollow cylinder of normal metal or insulator, rotating in a clockwise direction, thus constituting a circulating surface current density, $J$ Amps per unit length. As in a solenoid, this gives rise to a magnetic field $B_i$, pointing downwards inside the cylinder. In practice, we could make the cylinder the earthed plate of a cylindrical capacitor, with another normal metal cylinder outside, as shown in Fig. 1(b). The charges on this co-rotating outer cylinder make an equal and opposite contribution to the magnetic field, which changes direction but not magnitude and it now appears as $B_o$ in the gap between the two cylinders and is zero inside the middle cylinder. The charges on the opposite plates of a capacitor are equal and opposite, so the currents $J_o$ \& $J$ are equal and opposite in this case.

\begin{figure} [h]
\includegraphics[scale=0.5, bb = -50  0 300 480]{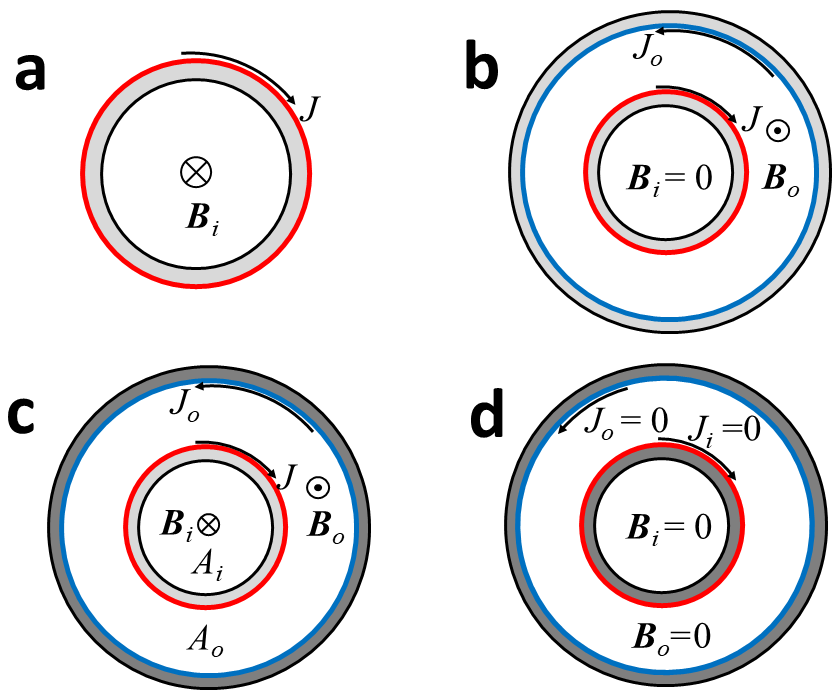}
\caption{\label{Fig 1} Evolution of a conceptual design. The colours on the surfaces of the components represent the relevant surface charges (red positive; blue negative). Normal metal is light grey, superconductor is dark grey. Surface current densities are $J$'s, magnetic fields are $B$'s and cross-sectional areas between the cylinders are $A$'s  (a) Isolated positively charged rotating normal metal cylinder; (b) with oppositely charged counter-electrode; (c) outer cylinder made of superconductor; (d) inner and outer cylinder of superconductor: this last change removes the magnetic field we intended to measure.}
\end{figure}
Recognising that this field is very small, the outer cylinder is then made of a superconductor to shield the interior from external magnetic fields, as in Fig. 1(c). Ignoring for now an effect called the `London moment', which we consider later, the total flux inside a superconducting cylinder is quantized, and taking it as zero before the voltage was applied to the capacitor, it remains zero afterwards. Therefore, a current density $J_o$ is induced on the inside of the outer cylinder, which gives rise to the field $B_o$ which maintains zero total flux inside that cylinder. $J_o$ is a dissipation-free current density, which may be partly due to rotating charges and partly supercurrent: its \emph{total} value is the only relevant quantity and is determined by flux quantization.  In this case of a superconducting outer cylinder, $J_o$ is not equal and opposite to $J$; these two current densities give rise to a magnetic field $B_i$ in the inner area $A_i$, plus an oppositely directed field $B_o$ in the outer area $A_o$, and zero total flux.

A normal metal inner cylinder will create Johnson noise in the magnetic field inside it, so one would like to make this superconducting too, as in Fig. 1(d). However, quantization means that the flux in the inner cylinder must remain zero as the capacitor is charged, and this means that the total current flowing around the inner cylinder, and the field inside it must be zero. Flux quantization in the outer cylinder ensures that the field in the outer area is also zero, so this modification has removed the effect we wish to observe! A way of avoiding this is shown in Fig. 2.
%
\begin{figure}
\includegraphics[scale=0.7, bb = -90 0 300 300]{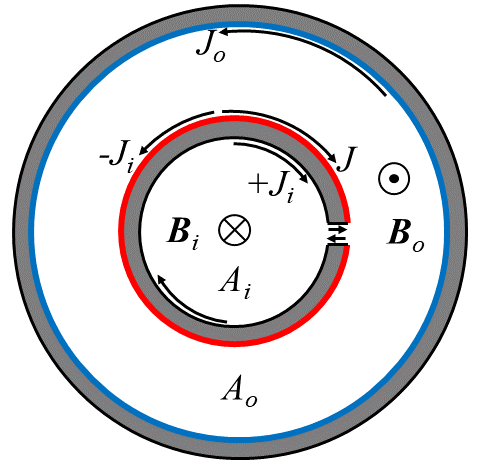}
\caption{\label{Fig 2 final} Schematic of an all-superconducting design. This represents the central cross-section of long concentric charged cylinders with the inner one having a narrow slit along its length. The slit forces the charge on it to rotate, again giving a surface current density $J$; additional supercurrents flow in the surfaces of the cylinders to maintain the magnetic field conditions required by superconductivity.}
\end{figure}

Here we have added a slit along the length of the superconducting inner cylinder. Electrostatics ensures that the charge remains on the outside of the inner cylinder and therefore must rotate with it. Flux quantization in the outer cylinder and zero magnetic field in the bulk of both superconductors are maintained by current densities $J_i$ and $J_o$. It will be noted that $J_i$ passes through the slit and flows on the inner surface of the inner cylinder, giving rise to the magnetic field $B_i$ inside it. This is the field we intend to measure: it is in an earthed, electric-field-free region. Our calculations will indicate\ that with $\pm \sim $ kV applied to an apparatus of radius $\sim 4$ cm, and a modern SQUID used for phase-sensitive detection of the signal, we could detect a signal due to the rotation of the Earth. With some scaling up, the signal would be large enough to give the rotation rate with good accuracy in a reasonable integration time. Furthermore, there are useful  experiments to be done on the electromagnetic effects of relative rotation of the cylinders and the detector, which can be performed with the small-scale apparatus and a rotation period of order 10 seconds, giving a much larger signal and hence a much shorter integration time.

Before proceeding to analyse this setup in detail, we make a few further comments:

(i) We would be using a superconducting SQUID to measure the very small fields produced in the apparatus. For the application envisaged - measuring the absolute rotation of the Earth - it is clear that the SQUID, like the rest of the Earth, will be rotating with the cylinders. This turns out to be important  because it is well-known that an uncharged rotating superconductor creates a tiny magnetic field throughout its bulk, known as the ‘London moment’~\cite{Brickman}. However if the SQUID \emph{is rotating with and inside a superconductor}, the London moment, is invisible, as the rotation affects the superconducting electrons in the SQUID in the same way as the bulk superconductor around it \cite{Brady2-online,LonMom}.  Hence, when the whole apparatus is in the rotating frame, we pick up only the effects of the rotating electrostatic charges. This was realized by Brady and is the basis of his work to develop an electromagnetic method of measuring rotation. This consideration is important because the field associated with the London moment can be comparable to that due to rotating charges if the SQUID is not co-rotating. We note that the London moment is a response due to the inertial mass of the electrons in the bulk uncharged superconductor, so does not change when the voltage applied between the cylinders is reversed. The London moment of a cylindrical superconducting shell has been considered~\cite{shell}. So long as the shell is thick compared with the microscopic superconducting coherence and magnetic penetration depths, and not close to $T_c$, then the bulk expression~\cite{Brickman} for the London moment applies.

(ii) At first sight, it may appear that the fields just inside and outside the slit in Fig. 2 cannot be different because there is no current flowing across the slit between them – apparently breaking Ampère’s theorem. However, near the slit, the electric field lines are no longer radial. As the split cylinder rotates, the change in the displacement of the electric field $d\textbf{D}/dt$, measured - like the current - in the non-rotating frame, gives rise to a tangential displacement current near the slit; this can be shown to `complete the circuit'. 
\section{Basic Theory}
\label{theory}
For simplicity, and to illustrate how the magnetic fields depend on the form of the apparatus sketched in Fig.~2, we carry out the calculations for the fields at the central cross-section of long cylinders, with the magnetic fields given by the expressions for the internal fields of long solenoids. Also, the currents due to charges will be carried within the electrostatic penetration depth of a metal surface and the supercurrents within the (longer, but still microscopic) magnetic penetration depth; we ignore this minor effect and take all currents to flow `in the surface'.

A charged cylinder of radius $r$ and length $\ell$, carries a surface charge density $\sigma = \varepsilon_0 E$ where $E$ is the radial electric field at the surface. (This still applies in the presence of a dielectric, where $\sigma$ is the sum of the mobile and polarisation charges, which rotate together in the apparatus. This point was also made in Ref.~\cite{Brady2-online}.) The total charge carried by the cylinder is $Q = 2\pi r \ell \varepsilon_0 E$. If the cylinder is rotating about its axis at angular frequency $\omega$, the circulating current per unit length, $J$ is given by:
\begin{equation}\label{JvsE}
 J=\frac{ \omega Q}{2 \pi \ell}=  \omega r \varepsilon _{0}E ~. 
\end{equation} 
This is the current density represented by $J$ in Figs. 1 $\&$ 2 and used in the theory below. We adopt the sign convention that clockwise currents and downward-directed magnetic fields are positive. We first calculate the cases in Fig 1(b) and (c), before turning to Fig 2.\par
In 1(b), the magnitudes of the charges on the normal metal outer and inner cylinders are equal and opposite and rotate together, so $J_o = - J$, and:
\begin{equation}
   B_{o}=- \mu _{0}J ~.  
\end{equation} 
Substituting for $J$, we obtain an expression for $B_o$ in terms of the electric field $E$ in the cylindrical capacitor.  
\begin{equation}
   B_{o}=- \mu _{0} \omega r \varepsilon _{0}E=- \omega rE/c^{2} ~.  
\end{equation} 
We note the term $c^2$ in the denominator, indicating that the magnetic field is small. 

We now treat the situation in Fig. 1(c) with an outer superconducting cylinder and an inner normal cylinder. Flux quantization gives an unchanged zero total magnetic flux inside the outer shield when the capacitor is charged: 
\begin{equation}\label{phi-zero}
   B_{o}A_{o}+B_{i}A_{i}=0~.   
\end{equation} 
The Meissner effect ensures that there will be zero field in the bulk of the superconducting outer cylinder and there is a magnetic field $B_o$ just inside it. Hence the net current density $J_o$ on its inside surface satisfies: 
\begin{equation}\label{Jo}
    \mu _{0}J_{o}=B_{o} ~.  
\end{equation} 
The current density $J$ carried around by the charge on the inner cylinder gives the difference between inner and outer fields: 
\begin{equation}\label{J-normal}
   B_{i}=B_{o}+ \mu _{0}J ~.
\end{equation} 
From Eqns.~\ref{J-normal}~$\&$~\ref{phi-zero}: 
\begin{equation}
   B_{i}= \mu _{0}J-B_{i}~. \left( A_{i}/A_{o} \right)~.     
\end{equation} 
Hence: 
\begin{equation}\label{Bi-normal}
   B_{i}= \mu _{0}J~.~A_{o}/ \left[ A_{i}+  A_{o}  \right] ~. 
\end{equation} 
Now we consider the intended design shown in Fig. 2, with a superconducting   split inner cylinder. Eqns.~\ref{phi-zero}~$\&$~\ref{Jo} still apply. The field $B_o$ in the outer area $A_o$ does not penetrate the bulk of the superconducting surfaces bounding it, so we have: 
\begin{equation}\label{Meiss-i}
    \mu _{0} \left( J-J_{i} \right) = -\mu_0J_0 = -B_{o} ~.  
\end{equation} 
Similarly, the field $B_i$ in the inner area $A_i$ is related to the surface current density on the inside of the inner cylinder: 
\begin{equation}\label{Meiss-h}
    \mu _{0}J_{i}=B_{i} ~.
\end{equation} 
From Eqns.~\ref{Meiss-h}~$\&$~\ref{phi-zero}: 
\begin{equation}
    \mu _{0}J_{i}=-B_{o}~. \left( A_{o}/A_{i} \right) ~.   
\end{equation} 
Hence, using Eqn.~\ref{Meiss-i}: 
\begin{equation}
    \mu _{0}J+B_{o}~. \left( A_{o}/A_{i} \right) =- B_{o}~.   
\end{equation} 
Thus: 
\begin{equation}
 B_{o}= -\mu _{0}J~.~A_{i}/ \left[ A_{i}+  A_{o}  \right] ~.  
\end{equation} 
Flux quantization (Eqn.~\ref{phi-zero}) relates $B_i$ to $B_o$, giving finally: 
\begin{equation}\label{B-inside}
   B_{i}= \mu _{0}J~.~A_{o}/ \left[ A_{i}+  A_{o}  \right]~.
\end{equation} 
This is exactly the same expression as Eqn.~\ref{Bi-normal}, the normal inner cylinder case. The reason is that the current density $J_i$ goes both clockwise and anticlockwise around the inner cylinder, and so does not alter the value of the field inside the inner cylinder relative to the normal case. However, as mentioned above, with a normal metal inner cylinder we have Johnson noise, whereas with superconducting parts we can employ further strategies described below to increase the sensitivity. These are necessary, as the fields produced are very small; for instance, with $\sim 1$ kV applied between cylinders of a few cm radius, and a gap between them of 0.2 mm, $B_i$ due to the Earth’s rotation is calculated to be $ \sim  10^{-16}$ T. Nevertheless, with a suitable design, fields of this magnitude can be detected with a SQUID magnetometer. 

In the following sections we consider three different designs to measure the rotation effect and analyze their sensitivity. In the case of the cylindrical designs, we assume `long' cylinders to obtain simple results. In practice, the outer cylinder would be longer than the inner cylinder with caps which mostly close the ends to provide a magnetic shield for the contents. In the gap due to the difference in length of outer and inner cylinders, the lines of flux due to $B_o$ would cross the ends of the inner cylinder and would become the equal return flux $B_i$. The ends of the inner cylinder would be partially closed in such a way as to allow axial magnetic flux to enter, but not radial electric field. In the Appendix, we also give an approximate treatment of Brady's second design\cite{Brady2-online} and compare its sensitivity with our designs.  
\section{Central pickup coil design}
\label{pickup}
We first define symbols for the relevant dimensions of the design represented schematically in Fig. 3. The outside radius of the inner cylinder is $r_o$, with a gap of $t$ across which the voltage is applied. The inside radius of the inner cylinder is $r_i$. We will find in a practical system, that the gap between outer and inner cylinders $t \ll r_o$, so that Eqn.~\ref{B-inside} may be written:
\begin{equation}\label{Bi}
   B_{i}= \mu _{0}J~.~2r_{o}t/ \left( r_i^2+2r_{o}t \right)~.
\end{equation}
\begin{figure}[h]
\includegraphics[scale=0.8, bb = -120 0 300 300]{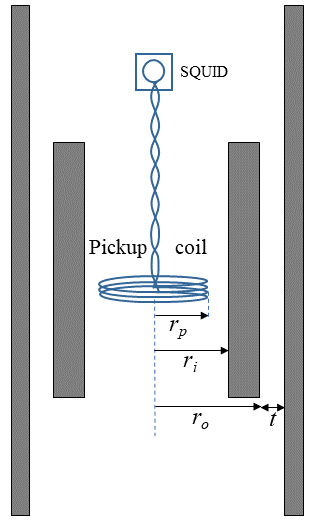}
\caption{\label{Fig 3} Schematic view of central pick-up design and SQUID detection system}
\end{figure}
At first sight, we would be inclined to make $r_i$ small in order to enhance $B_i$. As noted above, the fields are so small that a SQUID detector is required; this detects flux rather than field, so the criteria are different. The detection system is a pickup coil inside $r_i$ connected to the SQUID input coil, which feeds flux into the SQUID. In practice, the SQUID input coil would be in a screened region and arranged to make a negligible contribution to the field in the cylinders. 

The pickup coil of radius $r_p$ has $n$ closely-coupled turns, giving an inductance $L_p$. It responds to the flux turns $\Phi_p = n \pi r_p^2B_i$ to which it is exposed. This drives a current $I_S$ into the SQUID input coil of inductance $L_S$, which is tightly coupled to the SQUID. The SQUID and its integrated input coil is a readily available commercial item and is chosen to have a small equivalent input current noise $(S_I)^{0.5}$. In order to obtain maximum signal to noise in our proposed experiment we therefore need to maximise the current $I_S$ into the SQUID. Standard flux-transformer theory gives us:
\begin{equation}\label{Phi-t}
I_s = \Phi_p /(L_S+L_p) ~.
\end{equation}
If we treat the $n$ turns of wire on the pickup coil as close to each other, so that the coupling coefficient between turns $\approx$ unity and each turn has inductance $L_1 \sim \mu_0 r_p$, then $L_p = n^2 L_1$ and
\begin{equation}\label{n-turns}
   n= (L_p/L_1)^{0.5} = (L_p/\mu_0 r_p)^{0.5}~,
\end{equation}
$\Phi_p$ varies as $n$ but $L_p$ varies as $n^2$, so $I_S$ has a maximum when $L_p=L_S$ so
\begin{equation}\label{I_S}
    I_S = n B_i \pi {r_p}^2/2L_p = n B_i \pi {r_p}^2/2L_S~,
\end{equation}
and with $n$ now set to $(L_S/L_1)^{0.5}$, we can also write:
\begin{equation}\label{I_S2}
    I_S = B_i \pi {r_p}^2/2(L_S L_1)^{0.5} .
\end{equation}
We will ignore any screening of $I_S$ by the surrounding superconductor. This will be true if $r_p$ is not too close to $r_i$. We let $r_p=\gamma r_i$, with $\gamma \sim 0.8$, so that the area of the pickup coil is not too large a fraction of that of the hole.

Using Eqns.~\ref{Bi} $\&$ \ref{I_S}, we find 
\begin{equation}
    I_S= \frac{(\mu_0/L_S)^{1/2} \pi t r_o (\gamma r_i)^{3/2}}{(r_i^2+2 t r_o)} \times J ~,
\end{equation}
This is maximised when
\begin{equation}\label{ri-opt}
     r_i =(6 t r_o)^{1/2},
\end{equation}
giving:%
\begin{equation}\label{optimised}
    I_S=1.51 (\mu_0 / L_S)^{1/2}(t r_o)^{3/4} \gamma^{3/2} \times J ~.
\end{equation}
%
%
Using Eqn.~\ref{JvsE}, we can also re-express Eqn.~\ref{optimised} in terms of the applied voltage $V = Et$ and the angular frequency of rotation $\omega$:
\begin{equation}\label{optimised_V}
    I_S=1.51 (\mu_0 L_S)^{-1/2} r_o^{7/4}  t^{-1/4}\gamma^{3/2} \times  V \omega / c^2 ~. 
\end{equation}
This emphasises that small $t$ increases the signal, but in practice there will be a lower limit on $t$ for given $V$, set by the breakdown electric field in the gap, which can be increased by filling the gap with a dielectric. We have already noted that the presence of dielectric does not affect our derivation. 

For comparison with other designs it is useful to express the results in terms of $L_1$, the inductance of a single turn of the pickup coil, the flux $\Phi_1$ picked up by a single turn, and the total flux $\Phi_S$ in the SQUID input coil.
\begin{equation}\label{puL1}
    L_1 \approx \mu_0 r_p = \mu_0 \gamma r_i = \mu_0 \gamma (6 t r_0)^{0.5}  ,
\end{equation}
\begin{equation}\label{puPh1}
    \Phi_1 = \pi r_p^2 B_i = \frac{_3}{^2} \gamma^2 \pi t r_o \times \mu_0 J =  \frac{_3}{^2} \gamma^2 \pi r_o^2  V \omega / c^2  ,
\end{equation}
\begin{equation}\label{puL1Phi1}
    I_S = n \times \Phi_1 /2 Ls = \Phi_1 / 2 (L_S L_1)^{0.5},  
\end{equation}
and:
\begin{equation}\label{puL1Phi1}
    \Phi_S = L_S I_S = n \times \Phi_1 / 2 = (L_S / L_1)^{0.5}\Phi_1/2~.  
\end{equation}

In section~\ref{sensitivity}, we consider the role of $\Phi_1$, $L_S$ and $L_1$ in determining the signal to noise achievable with this and the two other designs described in sections~\ref{solenoids}  $\&$~\ref{disk}.
\section{Solenoidal pickup  design}
\label{solenoids}
\begin{figure}
\includegraphics[scale=0.6, bb = -80 0 250 200]{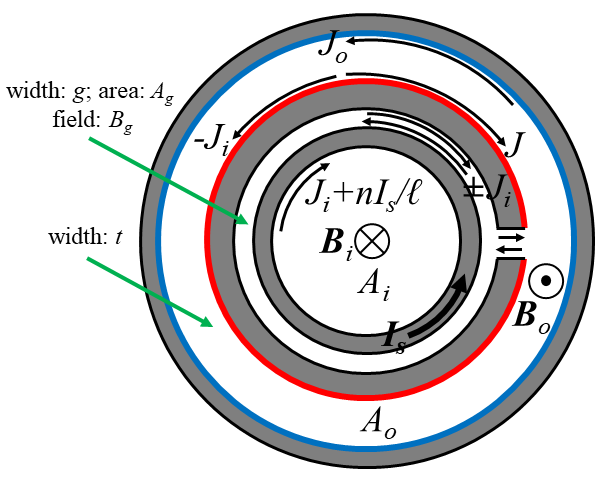}
\caption{\label{solenoid} Schematic diagram with a solenoidal pickup coil, which is represented in cross section by the inner grey circle. It is a superconducting cylinder of length $\ell$ cut into a single-layer solenoid of $n$ turns, with very small gaps between the turns. The ends of the solenoid are connected to a SQUID pickup coil, which takes a current $I_S$, so the current per unit length of the cylinder is $nI_S/ \ell$. The solenoid is fairly closely-fitting inside the split cylinder of inside radius $r_i$, with a gap of width $g$ and area $A_g = 2\pi r_i g$, kept small to reduce the flux in this region. The outer radius of the split cylinder is $r_o$ with a gap $t$ to the outer can, giving $A_o = 2 \pi r_o t$.  Note that main purpose of the split cylinder is to shield the SQUID circuit completely  from strong electric fields, and also from charging currents and their magnetic fields when the voltage is changed. This is achieved in practice by making the slit very narrow. }
\end{figure}

%
In the previous section, we considered a simple pickup coil. This would be suitable for in-principle tests for shorter rotation periods $\sim 10$ seconds, but we can greatly increase the sensitivity by having a pickup coil more strongly coupled to the inner split cylinder and occupying essentially all its area. The setup is shown in Fig.~\ref{solenoid}. As before, we carry out our calculations for the fields due to long cylinders, ignoring end effects. 
As in the situation described in Fig.~\ref{Fig 2 final}, the field $B_o$ in the outer area $A_o$ is related to $J$, $J_i$ \& $J_o$ by Eqn.~\ref{Meiss-i}. The field $B_g$ in the gap between solenoid and split cylinder is given by:
\begin{equation}\label{spi1}
   B_g = \mu _{0} J_i ~.  
\end{equation} 
The field in the central area is due to the sum of the induced current density $J_i$ on the inner surface of the split cylinder, plus the current per unit axial length due to the SQUID input current. (The latter is the sum of currents flowing on the inner and outer surfaces of the turns of the solenoid.). We therefore have:
\begin{equation}\label{spi0}
   B_{i}= \mu _{0}(J_i + n I_S / \ell)  ~.
\end{equation} 
Flux quantization in the SQUID input circuit gives:
\begin{equation}\label{spi2}
L_S I_S + n B_i A_i = 0 ~.  
\end{equation} 
Combining this with Eqn.~\ref{spi0} we find:
\begin{equation}\label{spi3}
  n \mu_0 J_i = -n^2\mu_0 I_S / \ell - L_S I_S / A_i~.  
\end{equation} 
Flux quantization in the outer cylinder gives: 
\begin{equation}\label{spi4}
   B_o A_o + B_g A_g + B_i A_i = 0~,   
\end{equation} 
and using Eqns.~\ref{Meiss-i}, \ref{spi1}~$\&$~\ref{spi2}:
\begin{equation}
    n \mu _0(J_i - J)A_o + n \mu _0 J_i A_g - L_S I_S = 0~.   
\end{equation} 
We then substitute for $J_i$, using Eqn.~\ref{spi3} to give a relationship between the SQUID current and the current density $J$ due to rotation:%
\begin{equation}
     n^2 \frac{\mu_0 A_o}{\ell}I_S +  n^2 \frac{\mu_0 A_g}{\ell}I_S + n \mu_0 A_o J  + L_S I_S \frac{(A_o+A_g)}{A_i}  +L_S I_S = 0~. 
\end{equation} 
Writing $A_{tot} = A_o+A_g+A_i$, we have after some manipulation:
\begin{equation}\label{spifinal}
 I_S = -\left( \frac{A_i}{ A_{tot} } \right)\frac{n \mu_0 
 A_o}{ (L_S + n^2 L_1)} J~.  
\end{equation} 
The minus sign on the RHS merely indicates that $I_S$ is in the opposite direction to $J$. The second term on the denominator is the effective inductance of the $n$-turn solenoid, which is $L_{eff} = n^2 \times L_1$, the inductance for a single turn, which is given by: 
\begin{equation}
 L_1 = \left( \frac{A_i}{ A_{tot} } \right)\frac{\mu_0 (A_o + A_g)}{\ell}~.
\end{equation} 
The numerator in Eqn.~\ref{spifinal} varies as $n$, but the denominator contains $n^2$, so we may vary $n$ to maximise $|I_S|$ for a given value of $L_S$. This gives:
\begin{equation}\label{spinvalue}
n^2 = L_S / L_1~.    
\end{equation}
This gives the SQUID current as:
\begin{equation}\label{modI_S}
 |I_S| = \left( \frac{A_i}{ A_{tot} }\right)\frac{n \mu_0 A_o }{2L_S} J = \left( \frac{A_i}{ A_{tot} }\right)\frac { \mu_0 2\pi r_o t} {2(L_S L_1)^{0.5}} J~.    
\end{equation} 
There is a further optimisation to perform: as in the previous section, we take $A_i \gg (A_o+A_g) $ so that the prefactor $A_i/A_{tot} \approx 1$. However, the value of $(L_1)^{0.5}$ in the denominator also depends on this factor and on $r_i$. Plotting $|I_S|$ versus $r_i$ results in a broad peak, at a value of $r_i$ that can only be obtained numerically. However setting $r_i \approx (4 t r_o^2)^{1/3}$ gives a sensitivity within $1\%$ of optimum, and still gives the prefactor $\approx 1$. We shall take it as unity for simplicity.
We may now write various results in terms of the dimensions of the apparatus: 
\begin{equation}\label{L_1-sol}
 L_1 = \frac{\mu_0 2 \pi (r_o t + r_i g)}{\ell},~
\end{equation} 
and to get a simple expression, we set $g = t$, and write for the mean radius $r_m = (r_i + r_o)/2$ to give:
\begin{equation}
 L_1 = \frac{\mu_0 4 \pi r_m t }{\ell}~.
\end{equation} 
We note that the value of $L_1$ is not proportional to the area $A_i$ inside the solenoid, but instead depends on the much smaller area of the gaps outside its windings. This is a consequence of the shielding effects of the surrounding superconductors.

 For a given applied voltage $V$, $J \propto E = V/t$, so the $t$ in the numerator of Eqn.~\ref{modI_S} disappears. We may also incorporate the expressions for $J$ and $L_1$ to give:
\begin{equation}\label{spiopt2}
 |I_S| \approx \frac {\pi \mu_0 r_0^2 } {(L_S L_1)^{0.5}} \times \varepsilon_0 V \omega = \left( \frac {\pi \ell r_o^4} {4 t r_m \mu_0 L_S } \right)^{0.5} \times V \omega / c^2~.
\end{equation} 
Another way of writing this is in terms of the flux transferred to the SQUID input inductance:
\begin{equation}\label{spiopt3}
 \Phi_S = L_S |I_S| \approx \left(\frac {L_S} {L_1}\right)^{0.5}\pi r_0^2  V \omega /c^2 = n \times \pi r_0^2  V \omega /c^2 = n \times \Phi_1 / 2~, 
\end{equation} 
where $\Phi_1 = 2 \pi r_0^2V\omega/c^2$ is the flux picked up by a single turn.

In section~\ref{sensitivity}, we shall consider what sensitivity this and other designs can give, and the role of the values of the flux and the inductances.
\section{Disk design}
\label{disk}
\begin{figure}[h]
\includegraphics[scale=0.6, bb = -90 0 350 500]{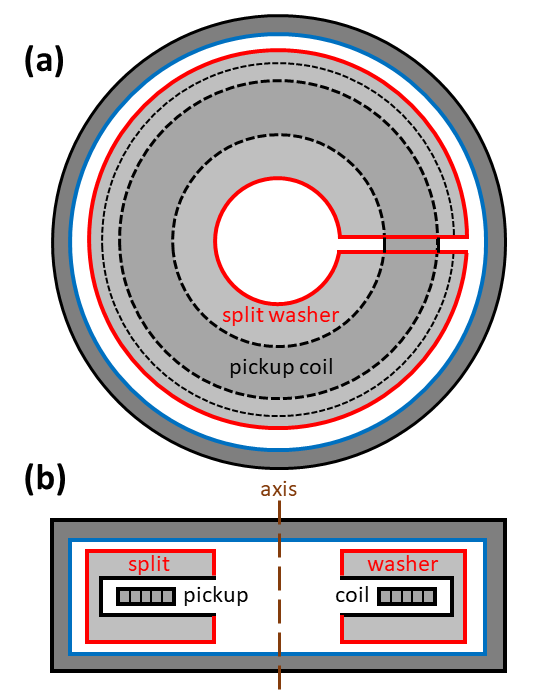}
\caption{\label{overall}Schematic drawing  of `disk' design;  (a) View along axis; (b) View in cross section. The $n$-turn pickup coil, split washer and outer shield are all superconducting. }
\end{figure}
\begin{figure}[h]
\includegraphics[scale=0.6, bb = -10 0 350 400]{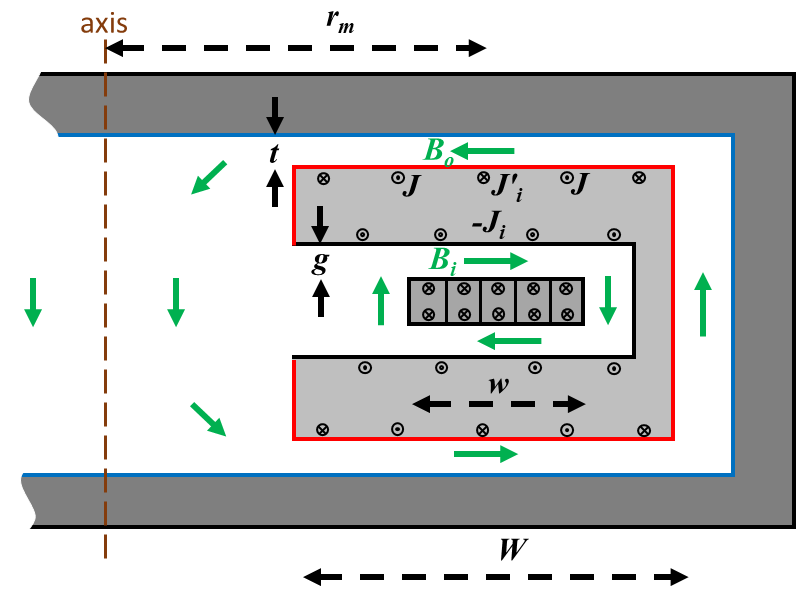}
\caption{\label{closeup} Detail of one side of the washer, showing the directions of surface current densities, magnetic fields and symbols for dimensions}
\end{figure}
In the cylindrical design of section~\ref{solenoids}, all the turns of the pickup coil share the same flux, and the inductance varies as the square of the number of turns. We now describe an alternative design, which consists of a number of identical cells which are magnetically isolated from each other, so that the total inductance of the pickup coils varies linearly with the number of cells, which are distributed along the rotation axis. A single cell is represented in Fig.~\ref{overall}, with detail given in Fig.~\ref{closeup}. This design is equivalent to the previous solenoid design, where the outer cylinder, which provided shielding and maintained flux quantization is now represented by the outer box. Inside, the slotted cylinder is folded into a C-cross-section slotted washer, which shields the `pancake' SQUID pickup coil. The coil has $n$ turns carrying a current $I_S$ within a radial width $w$. The spacings $g$ $\&$ $t$ and of the slot, and the axial length of the disk, would all be relatively much smaller than represented in the schematic figures.The algebraic analysis given below was verified using finite-element analysis modelling, to choose the dimensions finally adopted. 

The rotating charge on the outer surfaces of the washer gives rise to a current density $J$ on its top and bottom surfaces. The current in the SQUID pickup coil flows primarily in its top and bottom surfaces, giving rise to a surface current density $nI_S / 2w$. This induces a current density $J_i$ on the inner surfaces of the washer, which returns via the slot as a surface current density $J_i'$ on the outer surfaces. Most of this current flows on the surfaces of width $w$ $\&$ $W$, so continuity of current gives:
\begin{equation}\label{disk1}
 J'_i = J_i w/ W~.
\end{equation} 
All these currents give rise to magnetic fields $B_o$ $\&$ $B_i$, outside and inside the washer body. As before, the charged outer superconducting shield carries a surface current density between $B_o$ in the gap and zero field in the bulk of the shield. 

In this design, $B_o$ $\&$ $B_i$ are radial fields, so conservation of flux implies that in the regions of constant height $t$ $\&$ $g$ they should vary as $1/r$. Hence the surface current densities in the adjacent superconductors must also vary similarly with $r$. However for the purposes of analysis, we make the approximation of calculating the values that currents, fields and fluxes take at the mean radius from the axis, $r_m$. At this radius, 
\begin{equation}\label{disk2}
 B_o = \mu_0 (J'_i - J) =  \mu_0 (J_i w/W - J)~,  
\end{equation} 
and the total flux $\Phi_{out}$ passing through the axial hole due to $B_o$ is given by:
\begin{equation}\label{disk3}
 \Phi_{out} = 2\pi r_m t B_o = 2\pi r_m t \mu_0 (J_i w/W - J)~.
\end{equation} 
The field around the pickup coil obeys
\begin{equation}\label{disk4}
 B_i = \mu_0 J_i =  \mu_0 n I_S /2w~.  
\end{equation} 
The total flux $\Phi_{in}$ going through the pickup coil due to this current density is:
\begin{equation}\label{disk5}
 \Phi_{in} = 2\pi r_m g B_i = 2\pi r_m g \mu_0  J_i~.
\end{equation} 
Flux quantization in the SQUID input circuit gives:
\begin{equation}\label{disk6}
 L_s I_S + n(\Phi_{out} + \Phi_{in}) = 0~, 
\end{equation}
i.e.:
\begin{equation}\label{disk7}
 L_s I_S + n\left[2\pi r_m t \mu_0 ( nI_s/2W - J) + 2\pi r_m g \mu_0  nI_S/2w \right] = 0~. 
\end{equation}
Thus:
\begin{equation}\label{disk8}
 I_S \left[ L_s   + n^2\pi \mu_0 r_m ( t  /W + g/w) \right] = n2\pi r_m t \mu_0  J~. 
\end{equation}
The second term on the LHS is the effective screened inductance $L_c$ of the SQUID pickup coil in a single cell. Thus we may write the following expression for $I_S$ in terms of the flux $\Phi_c$ contributed by a single cell to the SQUID input:
\begin{equation}\label{disk9}
  I_S = \frac{\Phi_c }{( L_S   + L_c)}~. 
\end{equation}
Here:
\begin{equation}\label{disk10}
  L_c = n^2\mu_0 \pi r_m ( t  /W + g/w)~, 
\end{equation}
and
\begin{equation}\label{disk11}
\Phi_c = n2\pi r_m t \mu_0  J =  n \times 2\pi r_m^2 V \omega / c^2 ~, \end{equation}
while the flux in the SQUID input inductance may be written:
\begin{equation}\label{disk12}
  \Phi_S = L_S I_S = \frac{L_S}{( L_S   + L_c)} \Phi_c ~. 
\end{equation}
In the following section, we shall consider how the signal to noise may be optimised for a set of $N$ cells of this design and compare it with other designs.
\section{Optimisation of sensitivity and calculation of integration times to acquire a given accuracy}
\label{sensitivity}
We first consider designs, such as those described in sections~\ref{pickup} $\&$~\ref{solenoids}, in which we have a pickup loop with an effective inductance having the value $L_1$ for one turn. With $n$ closely-coupled turns it would have an inductance $L_1n^2$, and the $n$ turns would pick up a flux $n\Phi_1$. In this case, the current $I_S$ flowing in the SQUID input coil of inductance $L_S$  is given by:
\begin{equation}\label{sen1}
 I_S = \frac{n \Phi_1 }{ (L_S + L_1 n^2)} ~.  \end{equation}
A commercial SQUID (e.g. Ref.~\cite{SQUID}) has a spectral density of flux noise which may be represented as the current noise in its input coil $(S_{I})^{1/2}$ Amp.Hz$^{-1/2}$. We wish to maximise the signal to noise ratio $I_S / (S_{I})^{1/2}$. Now SQUID manufacturers generally provide a range of models with different values of inductance of the input coil, each of which has a different mutual inductance coupling it to the same design of SQUID. In these circumstances, $(S_{I})^{1/2}$ is not a constant independent of $L_S$. Rather, for ideal coupling of the input coil to the SQUID, the quantity $\frac{1}{2}L_S S_I$ is constant as it is equal to $S_E$, the energy sensitivity of the SQUID. The coupling coefficient for modern SQUID designs is close to unity, so for simplicity we shall write $L_S S_I$ as $2S_E$. This has the units J/Hz - the same as Planck's constant. Using $S_E$ we may write:
\begin{equation}\label{sen2}
 \frac{I_S}{(S_{I})^{1/2}} = \frac{I_S}{(2S_{E}/L_S)^{1/2}} = \left(\frac{L_S}{2S_{E}}\right)^{1/2}\frac{n \Phi_1 }{ (L_S + L_1 n^2)} ~. \end{equation}.
Optimizing this with respect to $n$, we find $L_S = L_1n^2$, giving:
\begin{equation}\label{sen3}
 \frac{I_S}{(S_{I})^{1/2}} =  \left(\frac{L_S}{2S_{E}}\right)^{1/2}\left(\frac{L_S}{L_1}\right)^{1/2}\frac{\Phi_1 }{ 2L_S} = \frac{\Phi_1 }{ 2(2S_E L_1)^{0.5}}~. \end{equation}
We see that the signal to noise is independent of $L_S$, and we may choose a SQUID from a range with a good energy resolution and adjust $n$ to match the apparatus to the input inductance of that SQUID. Also, we find that the signal to noise is maximised when the inductance of a single turn of the pickup coil is as small as possible for a given flux coupling $\Phi_1$ to a single turn. If we compare these quantities for the small pickup coil \emph{versus} the solenoidal pickup coil, we see that the former has both a smaller $\Phi_1$ and a larger $L_1$, so will be less sensitive. Its only advantage is that it allows SQUID and charged cylinders to be rotated independently, allowing experiments on the effects of relative rotation. 

 The inverse of the expression in Eqn.~\ref{sen3} is the fractional error in the rotational angular frequency $\omega$. The time necessary to obtain unity signal to noise in the rotation frequency is therefore given by:
\begin{equation}\label{tau}
    \tau \sim \frac{S_I}{I_S^2} =  \frac{8 S_E L_1 }{\Phi_1^2}~,
\end{equation}
and for noise  $\sim 1\%$ of the  signal we require a $10^4 \times$ longer measuring time. We shall later use this equation to estimate the integration times for various setups and rotation rates.

We now consider the optimisation of a design, such as the disk design described in section~\ref{disk}, in which we have a set of $N$ cells, each with a pickup loop of effective inductance $L_c$ picking up a flux $\Phi_c$. These are added in series  causing a current $I_S$ to flow in the SQUID input coil of inductance $L_S$:
\begin{equation}\label{sen4}
 I_S = \frac{N \Phi_c }{ (L_S + N L_c)}~.  \end{equation}
In this case, the signal to noise ratio is given by:
\begin{equation}\label{sen5}
 \frac{I_S}{(S_{I})^{1/2}} = \frac{I_S}{(2S_{E}/L_S)^{1/2}} = \left(\frac{L_S}{2S_{E}}\right)^{1/2} \frac{N \Phi_c }{ (L_S + N L_c)} ~. \end{equation}
 
To increase the sensitivity, the number $N$ of cells is chosen to be as large as possible, subject to experimental constraints. We can optimize  with respect to $L_S$, choosing a SQUID input coil best matched to the apparatus. Alternatively, for a given $L_S$, we can optimise the number of turns in the pickup coil in each cell, which gives the same sensitivity. In either case, we have $L_S = N L_c$, giving:
\begin{equation}\label{sen6}
 \frac{I_S}{(S_{I})^{1/2}}   = \left(\frac{N L_c}{2S_{E}} \right)^{1/2} \frac{N \Phi_c }{ 2N L_c} = \frac{\Phi_c }{ 2(2S_E L_c / N)^{0.5}} ~.
\end{equation}
Comparing this with Eqn.~\ref{sen3}, we see that the design with the higher sensitivity depends on how $\Phi_c$ compares with $\Phi_1$,  and $L_c/N$ with $L_1$. To know the value of integration time for a measurement, which involves a choice of actual dimensions and commercially available SQUIDs, we carry out calculations which are reported in the following section.

Here, we have analysed designs motivated by Brady's original suggestion~\cite{Brady1-pub}, using coaxial cylinders and disks. His later design~\cite{Brady2-online} consisted of two oppositely-charged flexible superconducting sheets separated by insulating material, wound some 50 times around a central superconducting cylinder to form a `Swiss roll', with the ends of each sheet connected by superconducting joints. The  current in one of these joints ran through the input coil of a SQUID. In the Appendix, we supply an approximate analysis of this design, which gives a similar sensitivity to our best designs. However, we have not considered it as a candidate as it presents some notable practical difficulties, as recorded by Brady~\cite{Brady2-online}. These are that the SQUID input circuit is exposed to high electric fields, making it sensitive to charging and leakage currents. Also, the sharp ends of the thin sheets making the turns of the spiral give a tendency to electrical breakdown, which can destroy the SQUID. The spiral has a very low inductance - smaller than that of the connecting leads. This reduces its sensitivity relative to our designs, which have total inductances considerably larger than the connecting leads. Finally, the thin superconducting sheets are not conducive to mechanical stability, which is necessary to avoid signals due to background flux. It has also been argued~\cite{Clive} that because the electric field transforms into a magnetic field in the rotating frame, there is a strong cancellation of the signal in Brady's design.
\section{Discussion}
\label{discussion}
There has been some controversy concerning what magnetic field due to a rotating charged cylinder would be observed when rotating with the cylinder. One might think that when rotating with the cylinder, there is no circulating current and therefore no magnetic field. However, just as new `fictitious forces'  – Coriolis and  centrifugal – appear in mechanics in a rotating frame, one might expect that there can be `fictitious currents' in the electromagnetic case. There is also the question of which frame the observer is in; these matters are discussed in Ref.~\cite{Clive}. An experiment to test this with constant voltage applied can be performed by a back $\&$  forth rotation while observing the resulting oscillation in magnetic field strength. Note that such measurements at constant voltage would be measuring the effects of relative motion with respect to the lab frame rather than absolute rotation. The earth's rotation would only cause a very small and constant offset of magnetic field value. The measurements could be done (using the lower sensitivity pickup-coil setup - or the higher sensitivity solenoid pickup if carefully engineered to ensure free rotation, while retaining a small gap) in three ways: 

(i) with the whole apparatus rotated together as envisaged for detecting the absolute rotation rate of the Earth. 

(ii) with the SQUID and its pickup coil on the same axis as the cylinders, but at rest while the cylinders are rotated - or 

(iii) vice versa - with the  SQUID rotating and the cylinders at rest. 

In the latter two cases, at zero applied voltage the SQUID would detect the field due to the London moment either of the surrounding rotating superconductor or of its own rotation. The London field is given by $B_L = 2\omega m/e$, where $m$ and $e$ are the free electron mass and charge\cite{Brickman}. When the voltage is applied, the SQUID would detect the sum of $B_L$ and $B_i$. This experiment would provide a test that the $B$ field observed in a rotating frame is the same as would be calculated if the rotating apparatus were considered from an inertial frame. By operating at constant voltage and using back and forth rotation, we avoid effects of charging currents and complications in making electrical connections, when the two parts of the apparatus are rotating differently.
%

\begin{table}
\centering
\begin{tabular}{||c|c|c|c|c|c|c||}
\hline 
\hline
Conditions~$\Downarrow$& Size~$\Rightarrow$&Small&Scaled&Small&Small&Small\\
\hline
Method of&Oscn. of~$\Rightarrow$&$\pm V$&$\pm V$&$\pm \omega$&$\pm \omega$&$\pm V$\\
measurement&@ freq.~$\Rightarrow$& 0.1~Hz& 0.1~Hz& 0.1~Hz& 0.01~Hz& 0.1~Hz\\
\hline
\multicolumn{2}{||c|}{Rotation  frequency~$\Rightarrow$}&1/day&1/day&0.1~Hz&0.01~Hz&0.01~Hz\\
\hline
Apparatus~$\Downarrow$ &Noise~$\Rightarrow$&10\%&1\% &1\%&1\%&1\%\\
\hline
Solenoidal &Int. $\tau$~$\Rightarrow$&6.3 hr&1 hr&31 ms&31 s&3.1 s\\
pickup &$\ell$/cm~$\Rightarrow$& 15 & 72 & 15 & 15 & 15 \\
design &$r_o$/cm~$\Rightarrow$& 3.5 & 17 & 3.5 & 3.5 & 3.5 \\
~&Opt. $n$  ~$\Rightarrow$&67&68&67&67&67\\
\hline 
Stacked&Int. $\tau$~$\Rightarrow$&7.8 hr&1 hr&38 ms&38 s&3.8 s\\
disks &$\ell$/cm~$\Rightarrow$& 15 & 57 & 15 & 15 & 15 \\
design &$r_o$/cm~$\Rightarrow$& 3.5 & 13 & 3.5 & 3.5 & 3.5 \\
~&Opt. $n$  ~$\Rightarrow$&3.6&1.8&3.6&3.6&3.6\\
~& disks $N$~$\Rightarrow$&19&62&19&19&19\\
\hline 
Single disk&Int. $\tau$~$\Rightarrow$&146 hr&~&0.7 s&700 s &70 s\\
$r_o = 3.5$~cm&Opt. $n$  ~$\Rightarrow$&16&~&16&16&16\\
\hline
Small &Int. $\tau$~$\Rightarrow$&317 hr&1 hr&1.5 s&1500 s&150 s\\
pickup &$\ell$/cm~$\Rightarrow$& 15 & 290 & 15 & 15 & 15 \\
design &$r_o$/cm~$\Rightarrow$& 3.5 & 68 & 3.5 & 3.5 & 3.5 \\
~&Opt. $n$  ~$\Rightarrow$&17&7.9&17&17&17\\
\hline 
\hline 
\end{tabular}
\caption{\label{Table 1} Table of calculations of the performance of the three designs. All calculations were carried out for a voltage of 1000 V applied across a gap $t$ of 0.2 mm. Where applicable, the gap $g$ was also set to 0.2 mm. All other dimensions are those given below or scaled up from them. To calculate the integration times required, we took the specifications of a particular Magnicon\textregistered~\cite{SQUID} SQUID, which has an input inductance of $1.8~\mu$H and a noise at 0.1 Hz of $(S_{I})^{0.5} = 1.48$ pA / $\surd$Hz. For frequencies in this region, $S_{I}$ rises as $1/f$, so it matters at what frequency the signal is phase-sensitively detected. For the back-and-forth rotations, which are to measure the effects of rotations relative to the lab frame, we have assumed the total rotation is half a turn before reversing, so that the frequency at which the rotation rate is oscillated is equal to the rotation rate.
For the small pickup coil design, we take $r_o = 3.5$~cm, which from Eqn.~\ref{ri-opt} gives $r_i \sim 0.65$ cm and set $\gamma = 0.8$. 
For the small solenoidal pickup coil design, we take $r_o = 3.5$~cm, giving $r_i \sim 1$~cm, from the discussion before Eqn.~\ref{L_1-sol}. For the disk design, we take $r_m = 3.05$ cm, $W = 0.9$ cm and $w = 0.5$ cm. $N \sim 19$ cells occupy approximately the same outer radius and length as the solenoidal pickup design. We also calculate the performance of a single cell.}
\end{table}
In Table I, we show under what conditions the measurements described above could be performed. In all cases we propose that either the applied voltage or the rotation direction is reversed, and the resultant change in SQUID output is phase sensitively detected. This avoids zero errors, but the detection frequency, and hence the SQUID noise, is limited by the frequency at which these reversals can be made. For short rotation periods, one could use the small pickup coil, which could comparatively easily be rotated independently from the cylinders. However, Table I also shows that this setup is not sensitive enough for detection  of the Earth's rotation, which would require the solenoidal pickup coil or the stacked disk design. We find that with the small size apparatus, the sensitivities of these two designs are almost equal. They have similar values of $\Phi_1$, the flux per single turn of pickup. For the stacked disk design, the larger value of $L_1$, the inductance per single turn, is counteracted by the $\surd N$ increase in sensitivity for $N$ cells in series. 

For high accuracy measurement of the Earth's rotation with either of these designs, a scaled up apparatus is required. To maximise sensitivity, only the radial extent of the components and the axial length are increased. All gaps and (in the disk design) the thickness of the disks are left unscaled, which keeps the electric field and inductance values constant. We find that the strongest scaling dependence is in $\Phi_1$, which increases as the square of the radius. The value of $L_1$ is independent of scale for both designs. However, the disk design has a $\surd N$ increase in sensitivity, and the  inductance of $N$ disks can be adjusted to match the available SQUID input inductance. The other way of increasing the sensitivity is to increase the applied voltage $V$, and the gap in proportion, so that one stays at a fixed fraction of the breakdown field of the dielectric. This increases both $\Phi_1$ and $L_1$, but since the latter appears as a square root (for instance in Eqn.~\ref{sen3}) this increases the sensitivity as $\surd V$.

We note that the SQUID parameters used in the Table correspond to an energy sensitivity $S_E \approx 3000~h$. With a measurement frequency of 0.1 Hz, we are a factor 30 below the frequency of 3 Hz, at which $1/f$ noise becomes important for this SQUID. Thus at high frequencies, it has white noise with $S_E \approx 100~h$. Since the integration time is proportional to $S_E$, it is advantageous to have both white noise and $1/f$ noise at the measuring frequency as close as possible to the quantum limit $\sim h$ .

It should be noted that to carry out any of these experiments is not without experimental difficulties. At a rotation speed of 1 Hz, we can generate a flux of $\sim 0.2~\Phi_0$, which is much larger than the noise level of modern SQUIDs, which is measured in $\mu \Phi_0 / \surd$Hz. However the flux due to the Earth's field passing through the SQUID pickup coil would be $\sim 10^8~\Phi_0$, so a very low field environment and constancy of the residual flux is essential for success.

The size of the signal depends on the geometry of the apparatus and the magnitude of the charging voltage, so it is unlikely that we would be able to provide an absolute calibration for measuring the Earth’s rotation rate to high accuracy. Instead, to measure the Earth’s rotation rate, one would mount the cylinder parallel to the Earth’s axis and servo the cylinder rotation to counter that of the Earth and hence reduce the signal to zero. To check that one obtains zero signal at zero rotation rate, the apparatus could be operated at 45º latitude, so that one can rotate the axis of the cylinder from $\parallel$ to $\perp$ to the Earth’s axis. (In both $\parallel~\&~\perp$ cases, the cylinder axis would be at $45^{\circ}$ to local vertical.) One can also check that the signal is zero at zero applied voltage. 

The operation of our apparatus can provide a check on the sometimes disputed expressions for electromagnetism in rotating frames (In a frame co-rotating with the cylinders there is no current flowing, but there should still be the same magnetic field\cite{Clive}). We may compare our apparatus with fibre-optic and ring-laser gyroscopes~\cite{photonics,RLaser}, which have to be `tickled'  to avoid locking. When this is done, they have been used successfully to measure the Earth's rotation rate to a rather greater accuracy than we expect to achieve - e.g. Ref.~\cite{RLaser}. However, they test Mach’s principle using optical techniques while our apparatus is using very low frequency electromagnetism, and rotating- or vibrating-mass gyroscopes~\cite{mech} are testing Mach’s principle applied to mechanics. 
\section{Conclusions}
\label{conclusions}
We have described the principles of operation for a detector of absolute rotation, which relies on the properties of superconductors when charged to high voltages. It can also act as a test/confirmation of the laws of electromagnetism at low frequencies in a rotating frame, and these measurements may be performed in a small-scale version of the apparatus. We have demonstrated that if the size of the apparatus is scaled up by a factor $\sim 5$ it should have sufficient sensitivity to measure the Earth's rotation to high accuracy and with a reasonable noise integration time. There are however various experimental hurdles that are likely to be encountered in its practical implementation.

\section{Acknowledgements}
\label{Acks}
We thank Antonello Ortolan, Giovanni Carugno \& Giuseppe Ruoso of the University of Padua for discussions and suggestions. The designs described in this paper are being implemented in a project at INFN Legnaro, Padua, Italy.

This research did not receive any specific grant from funding agencies in the public, commercial, or
not-for-profit sectors.
%






\appendix

\section{The Brady spiral}
\label{brady}
\begin{figure}
\includegraphics[scale=0.5, bb = -180 0 400 400] {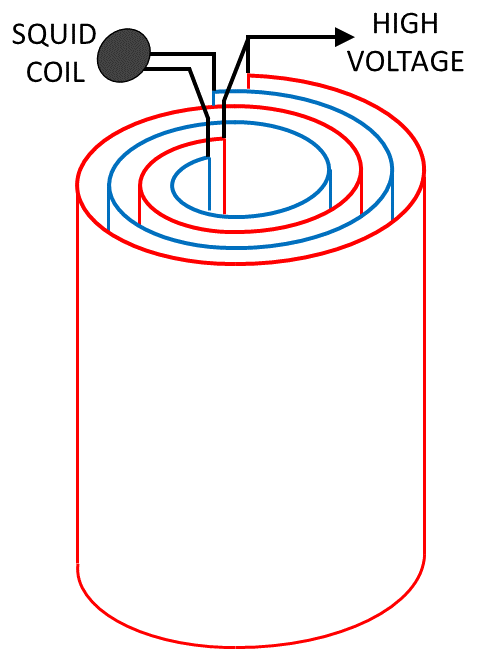}
\caption{\label{Brady3d} Schematic representation of the superconducting sheets and connections of the Brady Spiral~\cite{Brady2-online} For clarity, the central cylinder and the outer can are not included. A cross section is shown in Fig.~\ref{FigSP} for the purpose of analysis.}
\end{figure}

In Ref.~\cite{Brady2-online}, Brady considers a multi-turn double spiral consisting of two superconducting sheets separated by insulating layers. The sheets have thicknesses greater than the penetration depth so that there is zero magnetic field in their bulk, and the supercurrents flowing on each side of a sheet do so independently. The spiral is constructed by rolling the two sheets around a brass cylinder coated with a superconducting lead film, to form a “Swiss roll” as represented in Fig.~\ref{Brady3d}.  The inner and outer ends of the outer sheet are connected with a superconducting wire and a SQUID superconducting input coil is connected between the inner and outer ends of the other sheet so that each sheet forms part of a superconducting loop. The double spiral is surrounded by an earthed superconducting can and the central cylinder is also earthed.  A high voltage is applied to the outer spiral, while the inner spiral (which is connected to the SQUID input coil) is held at earth potential. The potential difference induces opposite charge densities on oppositely facing surfaces of the spirals, and also the inner surface of the outer can.  The whole apparatus is rotated about the spiral axis, and the rotating charges constitute surface-current densities, as in the case of the rotating cylinders discussed in the body of the paper.

In Brady's account of his work\cite{Brady2-online}, a \emph{detailed derivation} of its sensitivity is not given. However, he gives the dimensions and a calculated value, $L_{eff} = 11$ nH, for the effective inductance of his spiral. In our derivation below, we obtain a formula for $L_{eff}$, and using his dimensions, we obtain the same value. His approach of calculating the $\frac{1}{2}LI^2$ energy in the pickup circuit and comparing it with the energy sensitivity of the SQUID to calculate integration times is equivalent to ours. We do not know exactly how he treated the effect of stray inductances, but subject to that, we believe that our treatment of Brady's design is at least qualitatively similar to his.

\begin{figure}[h]
\includegraphics[scale=0.6, bb = -60 0 400 300] {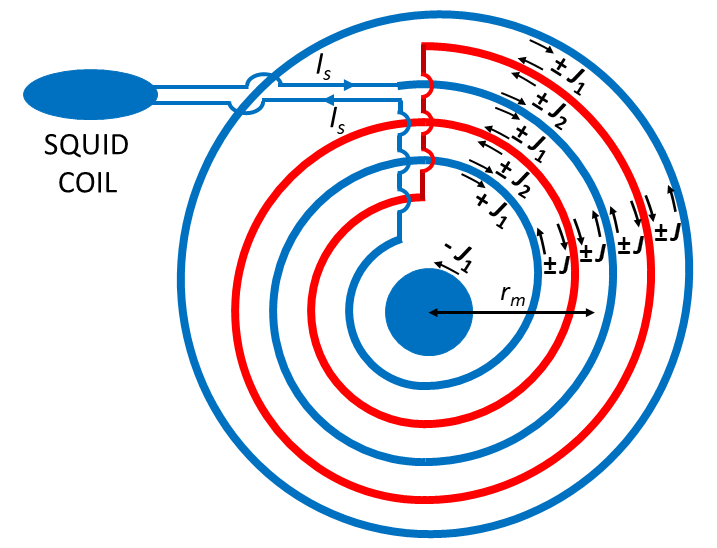}
\caption{\label{FigSP} The cross section of a 2-turn double spiral. (In practice, the spiral would have many turns.) The red and blue spirals represent positively and negatively charged superconducting sheets respectively and the blue circle represents a closed superconducting cylinder, which also carries charge. The opposite ends of each sheet are connected by superconducting wires as shown.
The arrows labelled $J$ represent the current densities resulting from the rotation; those labelled $J_1$ and $J_2$ represent induced supercurrent densities. The total current flowing through the blue link and the SQUID is $I_s$.
In the interests of clarity, the spirals are drawn with quite large spacings between them.  In practice, the spirals were actually wound quite tightly\cite{Brady2-online}.}
\end{figure}
We illustrate the principles of such a device, showing the double spiral in cross section in Fig.~ \ref{FigSP}. This apparatus can only be treated approximately, because the superconducting connections are made at one axial end of the spirals, so the induced current densities are non-uniform near the connections. However, for most turns of a multi-turn spiral, the currents are uniform. We note that, as the separation between the layers is much shorter than their total length, the magnetic fields in the gaps for these turns will be parallel to the axis of the spiral. However, there will be tangential components at the inner and outer ends of the spirals where the induced currents are non-uniform. The magnetic fields in the gaps are screened from the bulk of the superconducting sheets by surface currents and hence have the values $\mu_0(J-J_1)$ and $\mu_0(J-J_2)$. These fields point out of the page inside the charged red surfaces and into the page inside the charged blue surfaces. Assuming no trapped flux, the total flux within the outer can must be zero so we have
\begin {equation}
\mu_0 [ N(J - J_1)A -J_1A ] - \mu_0 N(J - J_2)A = 0,
\end {equation}
where we have generalised to a double spiral of $N$ turns ($N = 2$ in the Figures) and made the simplifying assumptions that all the relevant areas are equal to $A$ and all the turns have similar radii, so that the value of the rotation current $J$ is the same in each turn.  In the case of large $N$, it follows that $J_2 \approx J_1$, and for simplicity we shall take them as equal. The large $N$ approximation is in any case necessary, since this will minimise the additional effects on the flux from the non-uniform current densities at the ends of the spirals. Representing the axial length of the spirals as $\ell$, we see that the currents in the two external circuits are equal (and opposite) with the value $I_S = (J_1 + J_2) \ell = 2J_1 \ell$.
\begin{figure}[h]
\includegraphics[scale=0.7, bb = -20 0 500 300]{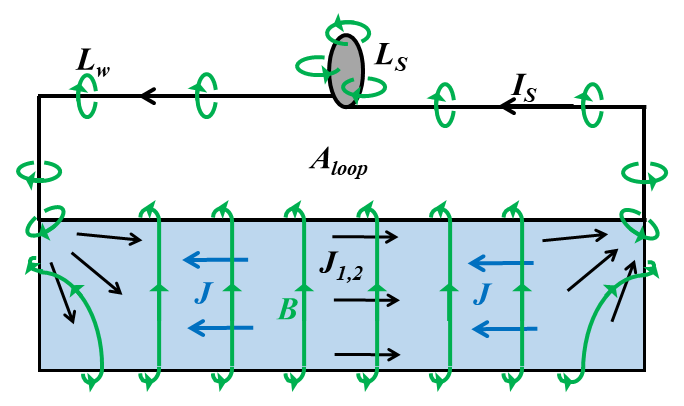}
\caption{\label{unwrapped} The blue spiral unwrapped: this shows the rotation current density $J$, which is nearly uniform, and the induced supercurrent densities $J_1$ \& $J_2$, which together form the current $I_S$ passing through the SQUID input coil. The total flux in this superconducting circuit is zero, so $L_SI_S$ plus the contribution of the stray inductance $L_wI_S$ is equal and opposite to that passing through the circuit area $A_{loop}$ due to the effects of $J$, $J_{1}$ \& $J_{2}$. The major contribution to this flux is that passing up one side of the sheet and down the other side. This flux only extends a distance $\sim t$ above the end of the spiral. This picture is only approximate, as there may be contributions to the flux from more distant turns of the spiral. These cannot be calculated in general, because their contribution depends on the distance of the upper part of $A_{loop}$ from the top of the spiral. These contributions will be small if the distance is small compared with the radius of the spiral but large compared with $t$.}
\end{figure}
Since both red and blue circuits are superconducting, the flux inside each must remain constant when the apparatus is charged and set into rotation. This will be ensured by a complicated pattern of induced supercurrent flows very near the axial ends of the spirals next to the external circuits. These  currents could only be calculated by a detailed numerical simulation, which is beyond the scope of this paper. A representation of the currents in one sheet is shown in Fig.~\ref{unwrapped}.

To analyze this situation, we use the condition that the total flux in the shorted circuit must remain zero. This means that the superconducting sheets will respond in such a way that the flux coming up inside the red spiral will not cross the top of this sheet and contribute to flux in that circuit. Instead, this flux will cross the blue sheet and go back down the gap inside it (See Fig.~\ref{unwrapped}). In making a detailed calculation of the effects of flux quantization, we note that the effective inductance of the spirals is very small, so that we must include an inductance $L_w$ due to the connecting wires (which includes inductance due to the high supercurrent densities at the ends of the spirals). If we ignored this stray inductance, the flux crossing the red spiral would be zero. In practice, a small amount crosses the red circuit and flux quantization ensures that this flux balances $L_w I_S$ due to the stray inductance of that circuit. 

The remainder of the flux inside the red circuit crosses the blue circuit, and is balanced by the flux in the stray inductance and the SQUID input circuit. Flux quantization in the blue circuit gives:
\begin {equation}
\mu_0(J-J_1)NA -L_w I_S = I_S(L_S + L_w )~.
\end {equation}
Using $J_1 \ell = I_S /2$, we obtain
\begin {equation}
 I_S = \frac{\mu_0 N A J}{L_S + 2L_w + L_{eff}}~,
\end{equation}
where $L_{eff} = \mu_0 N A/2\ell$ is the effective inductance of the spiral as viewed from the SQUID connections. This result may also be written: 
\begin {equation}
 I_S = \frac{2 J \ell L_{eff}}{L_S + 2L_w + L_{eff}}~.
\end{equation}
In the absence of the stray inductance, the best signal to noise would be obtained, as usual, by setting $L_S=L_{eff}$, but in this case we have to set $L_S = L_{tot} = (2L_w + L_{eff})$. This results in the optimised value for $I_S$ being reduced to:
\begin {equation}
 I_S = J \ell \times \frac{L_{eff}}{L_{tot}}=  J \ell \times \frac{L_{eff}}{L_S}~.
\end{equation}
The electrostatic interaction between charges on the sheets ensures that the charges rotate with the spirals, so the surface-current density, $J$, due to rotation is the charge per turn multiplied by the rotation rate. Given the above simplifying assumptions, we have
\begin{equation}
J=\varepsilon_0 E r_m \omega~.
\end {equation}
where $E$ is the electric field produced by the applied voltage.
We note that $NA$ is proportional to the total cross-sectional area of the Swiss roll, so for given $E$ (usually the limiting factor, to avoid breakdown in the insulator), $I_s$ is independent of $N$ for a given apparatus dimensions. However, the gap $t$ between the turns is necessarily smaller at large $N$, so the required voltage $V$ decreases, which is an experimental advantage, although balanced by the fact that thinner turns are less dimensionally stable. We also note that the number of turns does not alter the effective inductance of a spiral. This inductance is smaller than the stray inductance, which reduces the sensitivity. The input inductances of modern SQUIDs can be as low as $L_{tot}$, so the additional complication of a flux transformer is not required.

An alternative expression for the optimised $I_S$ may be written in terms of the spacing $t$ between turns and the applied voltage $V$:
\begin {equation}\label{Isspiral}
 I_s = \frac{\mu_0 N \pi r_m t \times \varepsilon_0 (V/t) r_m \omega }{L_S} = \frac{ \pi N r_m^2 }{L_S}\times  V\omega / c^2~.
\end{equation}
where we have written $A = 2\pi r_m t$. This can be used in comparing sensitivities.

Throughout this discussion, we have assumed that the number of spiral turns, $N$, is large, which is certainly true in Brady’s set-up where $N = 54$.  The more general case would require numerical simulation as the induced supercurrents and fields are non-uniform. We have not attempted this, but believe that all corrections to our treatment are $\propto 1/N$. We have also used the simplification that all turns of the spiral have essentially the same radius; we expect that using a mean radius $r_m$ will be a good approximation. Assuming this, there is an optimum value for the inner radius $r_i$ of the spiral for given outer radius $r_o$. Reducing $r_i$ increases $N$, but reduces $r_m$ in Eqn.~\ref{Isspiral}. Numerical optimisation gives the optimum $r_i \approx 0.34 r_o$, so $r_m \approx 0.67r_o$, which is very similar to the optimum for our solenoid design.   

We now discuss the sensitivity of the spiral design. For comparison with our designs, we take a spiral with the same external dimensions, gaps and applied voltage as our solenoidal design with outside diameter 3.5 cm. Hence we take the width of the insulating gaps (and also the thickness of the superconducting lead sheets) as 0.2 mm. This gives 0.8 mm as the pitch of the double spiral. Fitting the spiral on an $r_i = 1.2$ cm inner cylinder inside an outer can at 3.5 cm radius gives $N=28$ turns within an outer radius of $r_o =3.4$ cm giving ($r_m = 2.3$ cm). The axial length of the spiral would be 15 cm. We assume that the stray inductance of both connecting wires is $\sim \mu_0 r_o$.
This gives the effective inductance of the spiral $L_{eff} = 3.4$ nH, and $2L_w = 85.5$ nH, so the output inductance of the spiral is $L_{tot} \sim 89$ nH. This is much less than $L_S =1.8~\mu$H, the input inductance of the SQUID we considered in the main text. The easiest way to deal with this is to assume that another model from the same range has $L_S = L_{tot} \sim 90$ nH and the same value of $S_E$: this would give the same signal to noise as if an ideal flux transformer were used to match $L_{tot}$ with $L_S =1.8 \mu$H.
Taking all this into account, we find that the ratio between signal to noise for the spiral and that for our solenoidal pickup model is:
\begin {equation}
 \frac{(I_S/(S_I)^{0.5})_{spiral}}{(I_S/(S_I)^{0.5})_{solenoidal}} = 2\frac{r_m^2 }{r_o^2} \left[\frac{L_{eff}}{L_{tot}} \right]^{0.5}N^{0.5} \sim 0.9~.
\end{equation}
 Thus we find that the multi-turn spiral has \emph{in principle} a similar sensitivity to our solenoidal pickup design. However, the practical problems found by Brady with this design, which are described in the main text, indicate that the designs described in this paper are preferable.
%










%

\end{document}